\documentclass[fleqn,12pt,twoside]{article}
\usepackage{espcrc1}

\usepackage{epsfig}

\newcounter{fig:asym}

\newcommand{\lsim}{\raisebox{-0.5mm}{$\stackrel{<}{\scriptstyle{\sim}}$}}

\newcommand{\qsq}{\mbox{$Q^2$}}
\newcommand{\pom}{I\!\!P}
\newcommand{\xpom}{x_{\pom}}

\newcommand{\pbinv}{\mbox{${\rm pb^{-1}}$}}
\newcommand{\fbinv}{\mbox{${\rm fb^{-1}}$}}

\title{\vspace{-1.0 cm}Deeply Virtual Compton Scattering at H1 and ZEUS\footnote{Contribution to the proceedings of QCD-N'02 Workshop -
        Ferrara (I), 3-6 april 2002}}                        

\author{Laurent Favart
        \address{I.I.H.E., Universit\'e Libre de Bruxelles, Belgium\\
        E-mail: lfavart@ulb.ac.be}}

\begin{document}

\maketitle

\begin{abstract}
Results on Deeply Virtual Compton Scattering
at HERA measured by the H1 and ZEUS Collaborations are presented.
The cross section, measured for the first time, is reported for
by H1 and ZEUS for $Q^2$ above a few GeV$^2$ in the low $x$ region.
The measured cross section is discussed and
compared to different predictions.
\end{abstract}

\section{INTRODUCTION}

\begin{figure}[htbp]
 \vspace{-1.2cm}
 \begin{center}
  \epsfig{figure=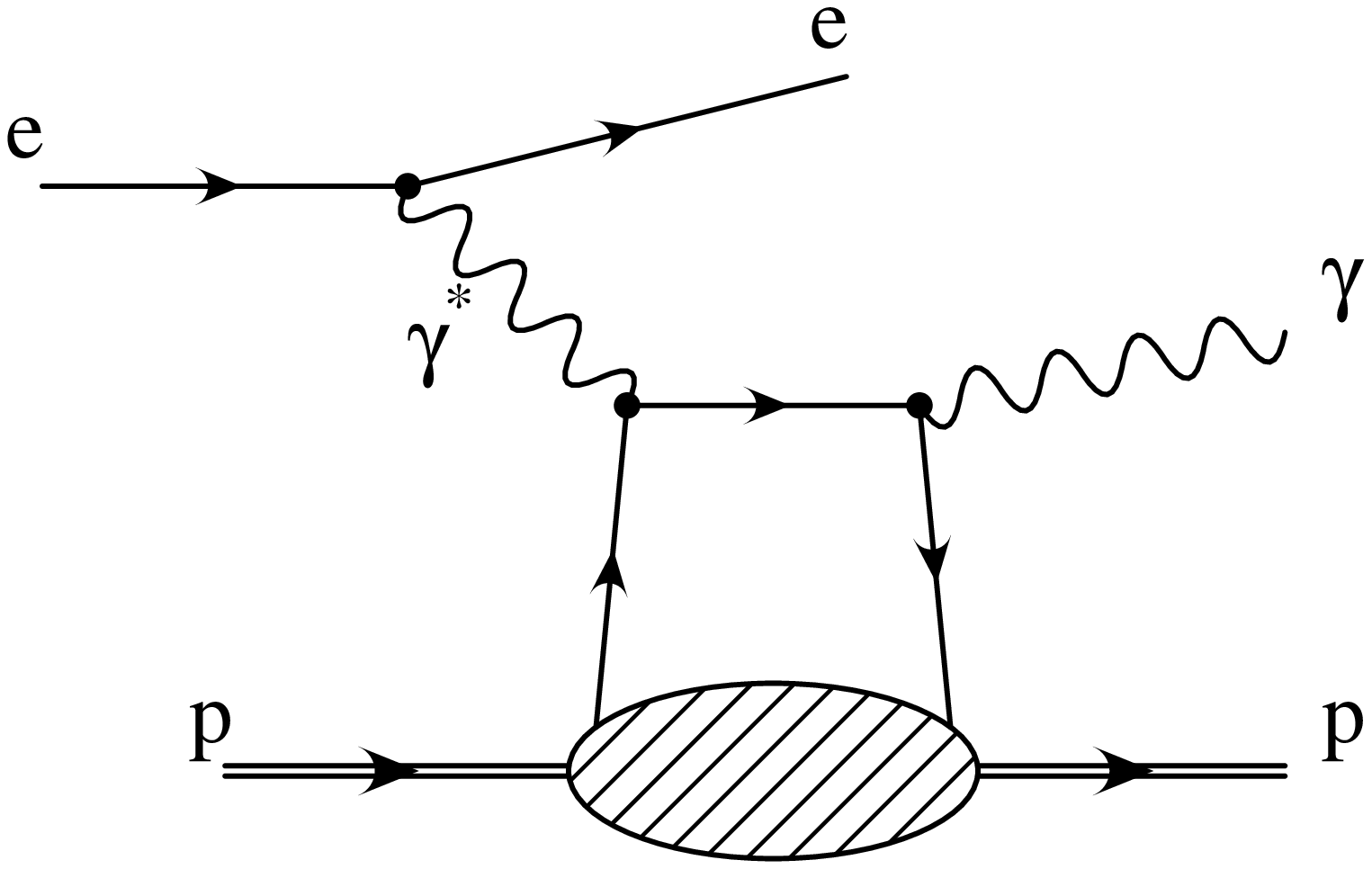,height=0.24\textwidth}
  \hspace*{0.5cm}
  \epsfig{figure=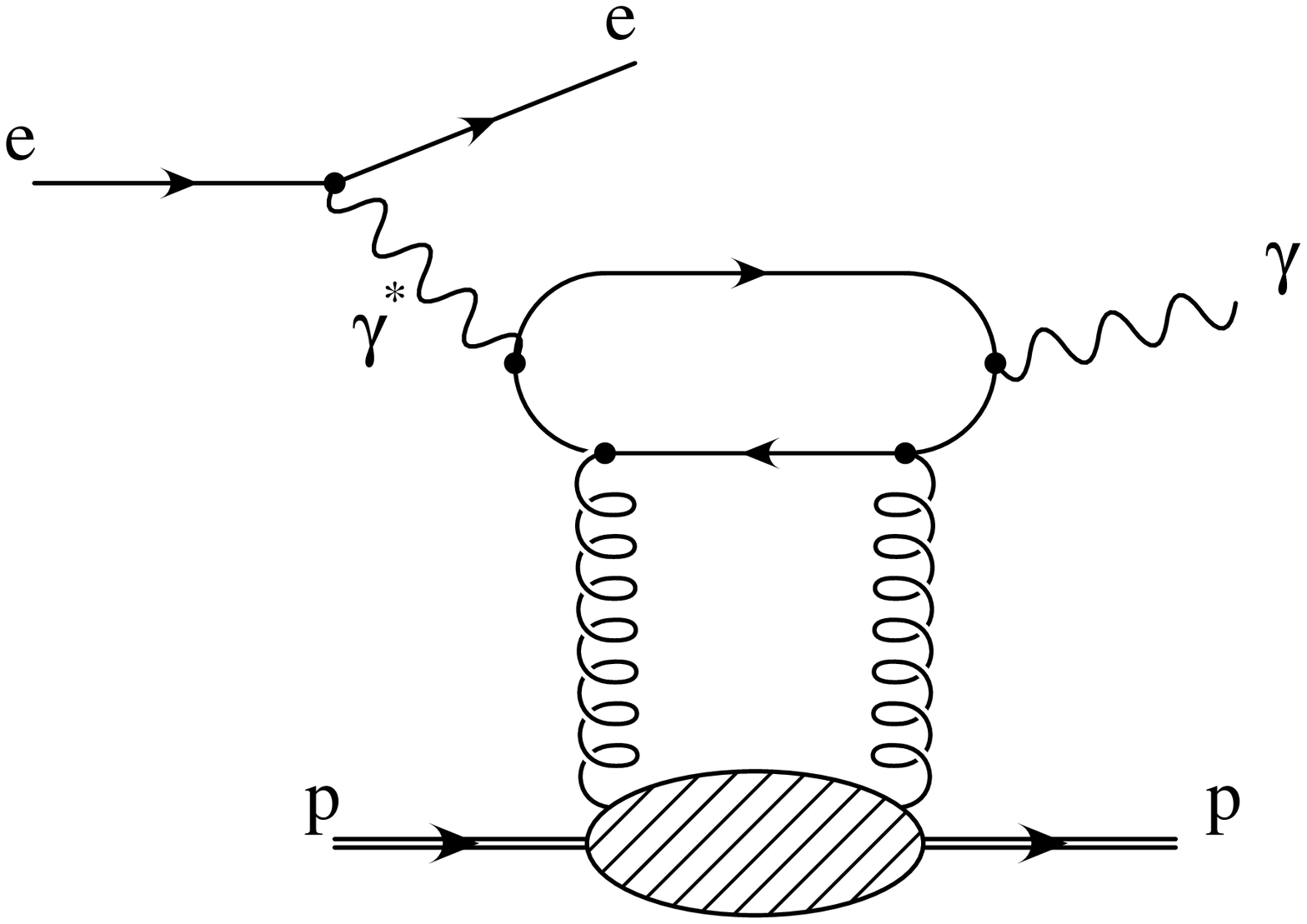,height=0.26\textwidth}
  \end{center}
 \vspace{-1.0cm}
 \caption{The DVCS \,\,\,{\bf a)} at LO \hspace{4.5cm} {\bf b)} at NLO.}
 \label{fig:diag1}
\end{figure}

 At the high energy of $\sqrt{s} \simeq 300$ GeV delivered by HERA using
colliding electron (27.5 GeV) and proton beams (820 GeV), the Deeply
Virtual Compton Scattering process (DVCS) 
$e p \rightarrow e \gamma p$
is of diffractive nature.
Comparing to the lower energy experiments CLAS\cite{Elouadriri} and 
HERMES\cite{Hasch,Ellinghaus}, additionally to the 
direct quark contribution (LO contribution shown in Fig.~\ref{fig:diag1}a),
the color singlet two gluon 
exchange is also expected to have a sizable contribution
(NLO - Fig.~\ref{fig:diag1}b).

A considerable interest of the DVCS comes from the particular access 
it gives to Generalised Parton Distributions (GPD) through the
interference term with the Bethe-Heitler process.
The high energy situation of H1 and ZEUS experiments give them the
unique opportunity to constrain the gluon contribution to GPDs and to
study the evolution in \qsq\ of the quark and gluon distributions.

Here we report the first cross section measurement of the
Deeply Virtual Compton Scattering, performed by the H1 and ZEUS experiments.
The cross section measurement are compared to the theoretical predictions
and future plans for the DVCS measurement at HERA are briefly presented.

\section{ANALYSIS STRATEGY}

\begin{figure}[htbp]
 \vspace{-0.8cm}
 \begin{center}
  \epsfig{figure=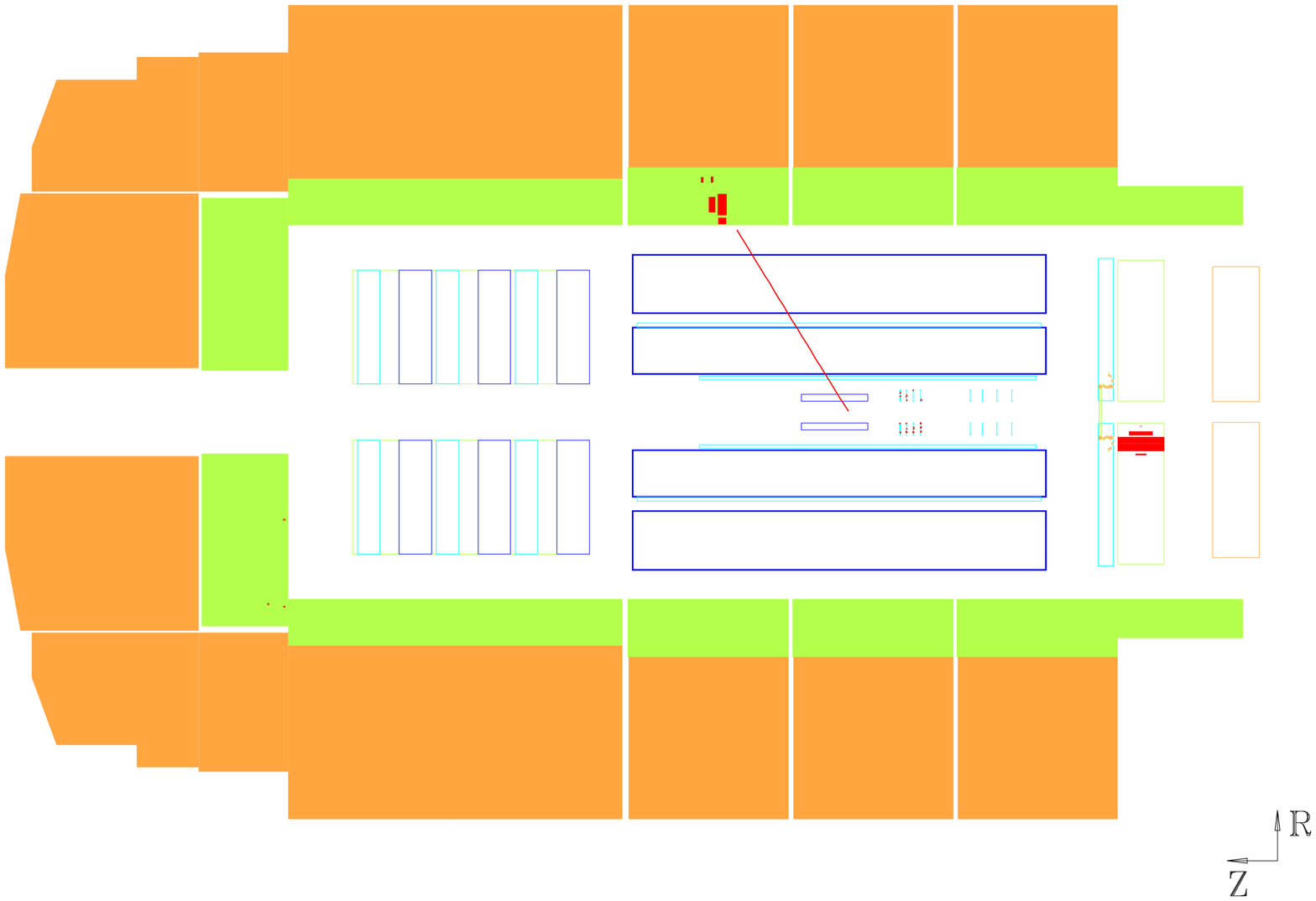,height=6.0cm,angle=90}
  \hspace{2.0cm}
  \epsfig{figure=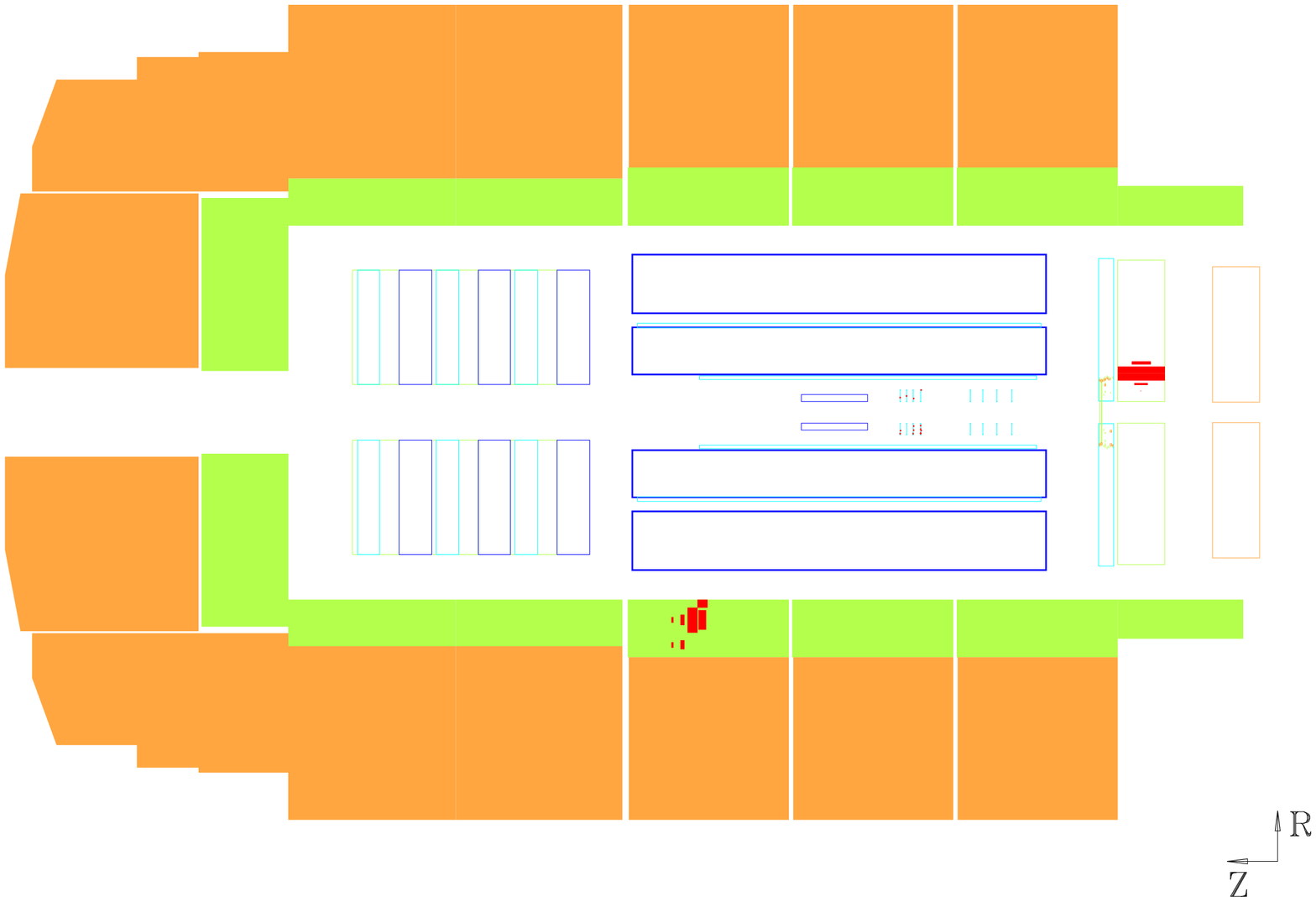,height=6.0cm,angle=90}
  \vspace{-1.0cm}
  \caption{{\bf a)} Control sample topology.
           \hspace{1.5cm}
           {\bf b)} Enriched DVCS sample topology.
   }
 \vspace{-1.0cm}
  \label{fig:conf}
 \end{center}
\end{figure}

Both H1 and ZEUS adopted the same analysis strategy. 
Around the interaction region, H1 and ZEUS 
are equipped with tracking devices surrounded by calorimeters.
Since the proton escapes the main detector through the beam pipe only 
the scattered electron and photon are measured. Therefore the
event selection is based on demanding two electromagnetic clusters,
one in the backward (i.e. the direction of the
incoming electron, $\theta=0$) and one in the central or the 
forward part of the detector ($\theta \lsim 140^o$).
If a track can be reconstructed it has to be
associated to one of the clusters and determines the electron candidate.
These events are then subdivided into two mutually exclusive samples.
The {\it control sample} (Fig.~\ref{fig:conf}.a) contains events with the
electron candidate detected in the central or the forward part of the
detector. This sample is largely dominated by Bethe--Heitler events,
although some additional backgrounds due to dilepton production by
photon-photon interaction and due to diffractive $\rho$ electroproduction have
to be considered~\cite{Rainer}. 
The DVCS contribution is kinematically suppressed by the large \qsq\ value.
The {\it enriched DVCS sample} (Fig.~\ref{fig:conf}.b), 
complementary to the control sample,
contains both DVCS and Bethe--Heitler events, and a small background 
contribution due to $\omega$ and $\phi$ diffractive 
electroproduction.
  
To enhance the DVCS contribution in comparison 
to the Bethe-Heitler process the phase space has to be restricted
by demanding the photon candidate in the
forward part of the detector.

The H1 analysis selects more specifically the elastic component by using, 
in addition, forward detectors, placed close to the 
beam pipe, are used to identify 
particles originating from proton dissociation processes. 

Both experiments compare their results to the prediction of
L.~L.~Frankfurt, A.~Freund and M.~Strikman (FFS)~\cite{FFS}
calculated in QCD at LO and leading twist. 
It is important to notice that, into this approximation,
the interference term cancels out when
integrating over the azimuth angle of the final state photon (as in the 
analysis here below). 
Therefore the pure DVCS cross section can be measured 
subtracting the Bethe-Heitler contribution to the enriched DVCS sample.


\section{ZEUS RESULTS}

\begin{figure}[h]
 \begin{center}
  \epsfig{figure=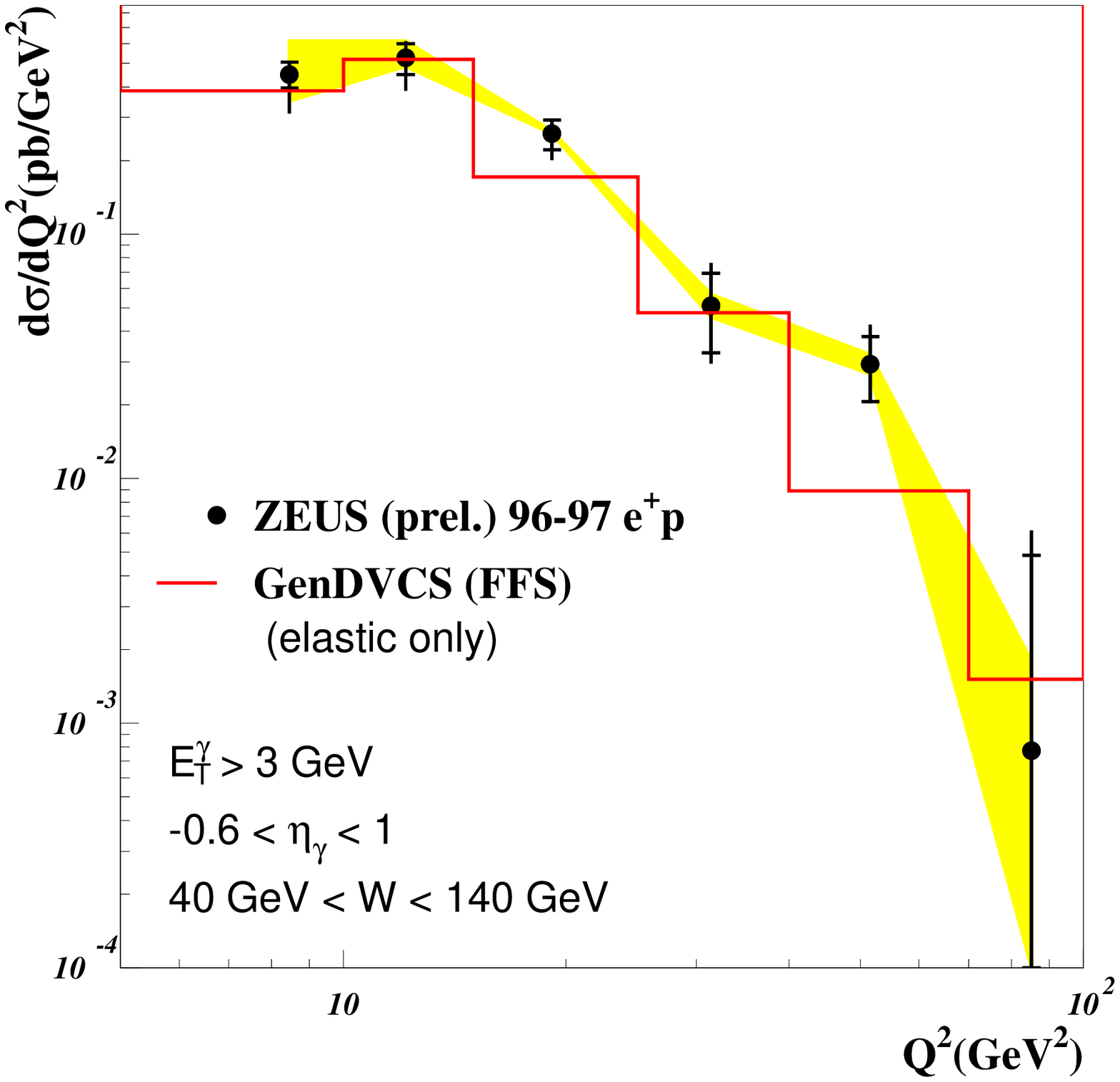,width=0.35\textwidth}
  \hspace{1.0cm}
  \epsfig{figure=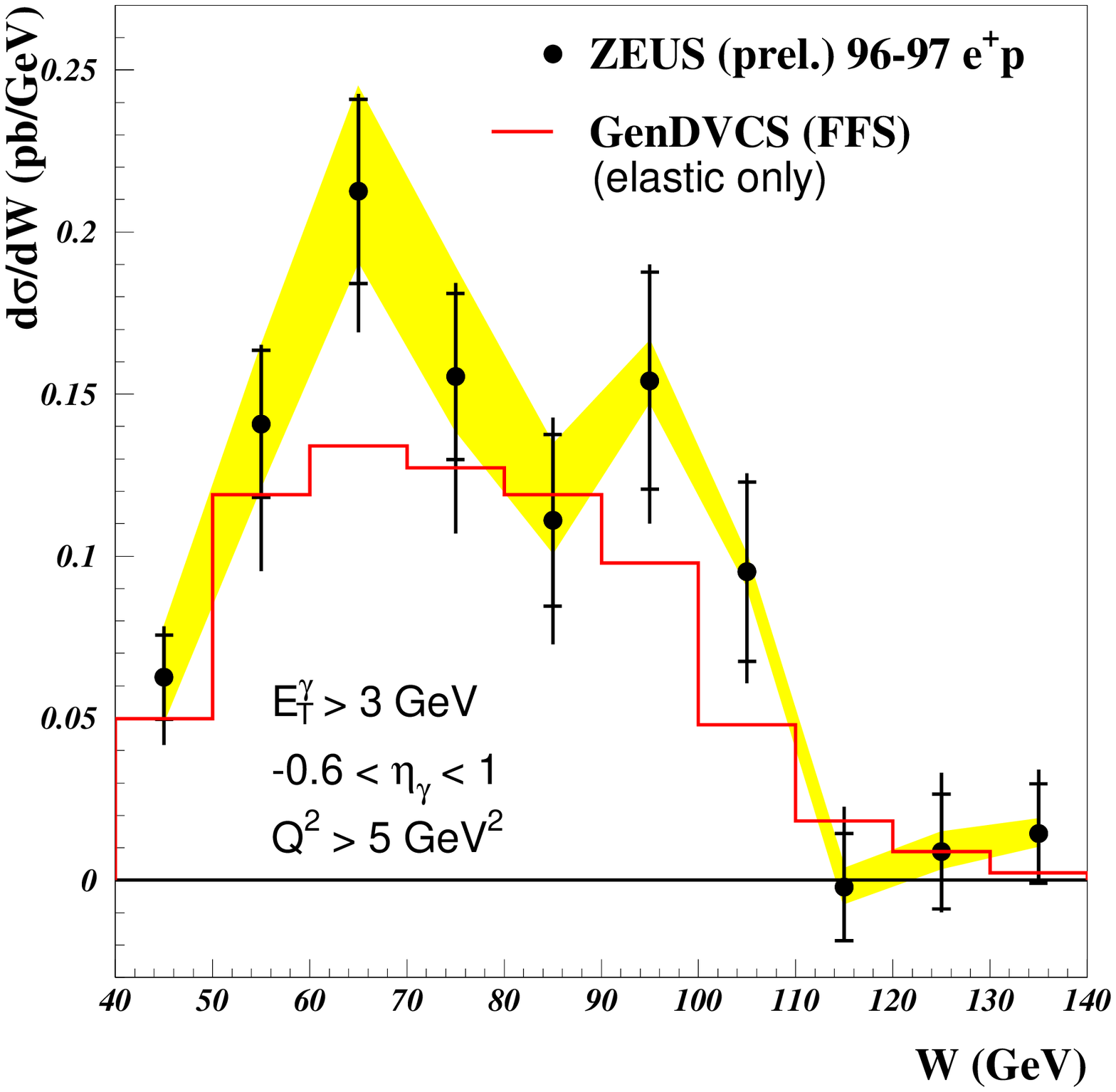,width=0.35\textwidth}
  \vspace{-0.6cm}
  \caption{
   DVCS contribution to the $e p \rightarrow e \gamma p$ cross section
   as a function of $Q^2$ (left) and of $W$ (right). The measurement is
   compared to the FFS prediction.
   }
  \label{fig:z_sig} 
  \vspace{-1.6cm}
 \end{center}
\end{figure}

This analysis~\cite{ZEUS-DVCS} is based on a sample corresponding 
to an integrated luminosity of 37~\pbinv. 
The backward electromagnetic cluster 
is required with an energy above 10 GeV, and the central or forward
cluster ($-0.6 < \eta < 1.0$) with more than 3 GeV.
The photon virtuality $Q^2 > 5\,{\rm GeV}^2$ is demanded, 
the $\gamma^*-p$ energy, $W$, is such that $40 < W < 140\,{\rm GeV}$.
%
%
The $e-p$ DVCS (preliminary) cross section is shown as a function 
of $Q^2$ and of $W$
in Fig.~\ref{fig:z_sig} and is compared to the FFS prediction.
The measurement is well described by the FFS prediction assuming a value of 
$b=4.5$ GeV$^{-2}$ for the exponential $t$ slope. It has to be noted that the
background from the proton dissociation (around 20\%) has not been
subtracted.
 
\section{H1 RESULTS}

In the H1 analysis~\cite{H1-DVCS}, 
corresponding to an integrated luminosity of 8~\pbinv, 
the DVCS cross section is measured in the kinematic region:
$2 < Q^2 < 20\,{\rm GeV}^2 $,
$ |t| < 1\,{\rm GeV}^2$
and
$30 < W < 120\,{\rm GeV}$.
The proton
dissociation background has been estimated to of 16\%~$\pm~8$\% and is 
subtracted 
statistically assuming the same $W$ and $Q^2$ dependence as for the elastic
component.

It may be worth to comment on possible $\pi^o$ contamination as suffer
the DVCS measurements of HERMES and CLAS. Due to the high energy regime, 
the direct $\pi^o$ production (via Regge pole exchange) is suppressed, 
and the possible $\pi^o$
contamination from tails of low multiplicity DIS events is suppressed by
the large rapidity gap requirement (see detailed study in~\cite{Rainer}).

\begin{figure}[htbp]
 \begin{center}
  \epsfig{figure=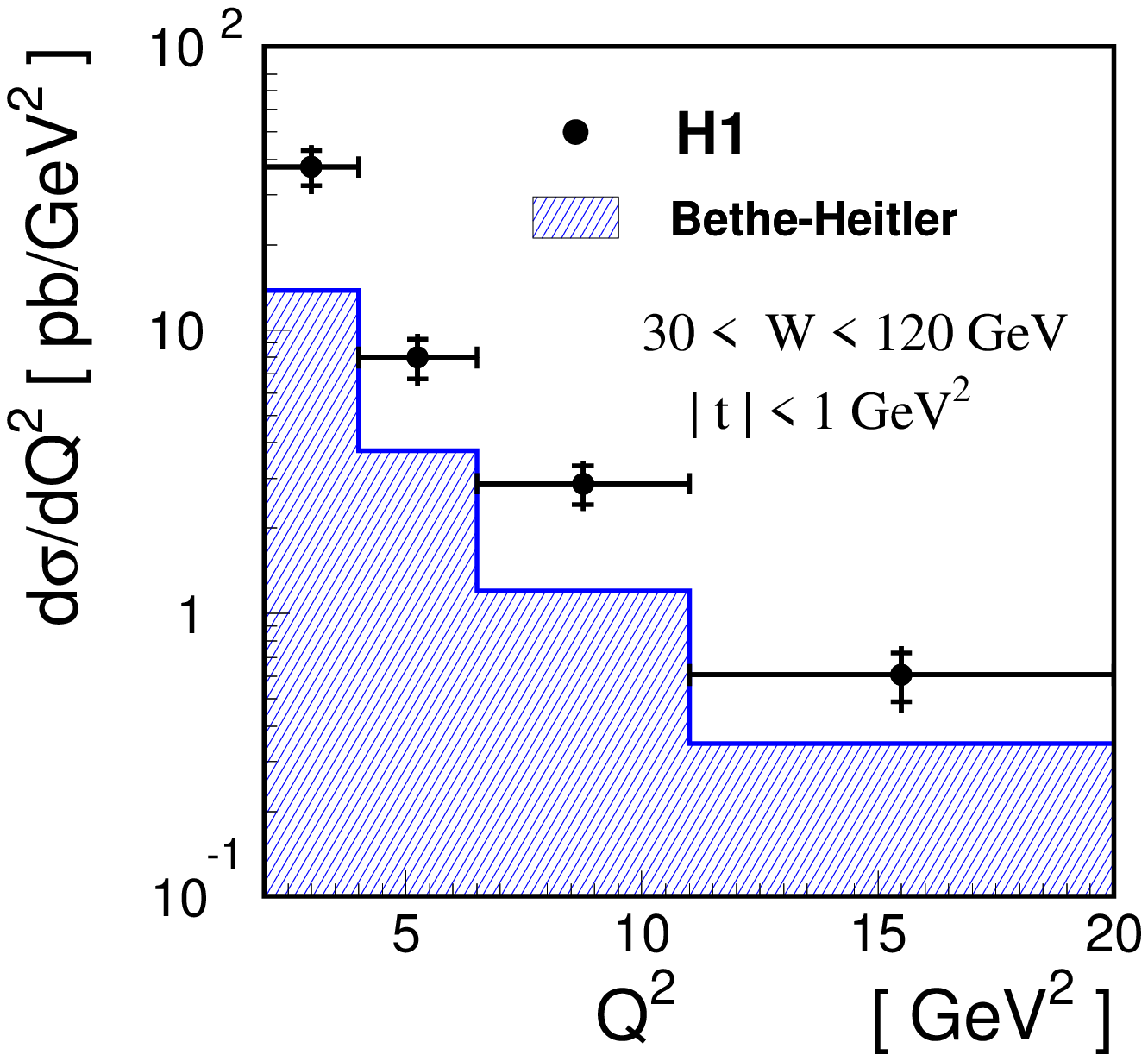,width=0.4\textwidth}
  \epsfig{figure=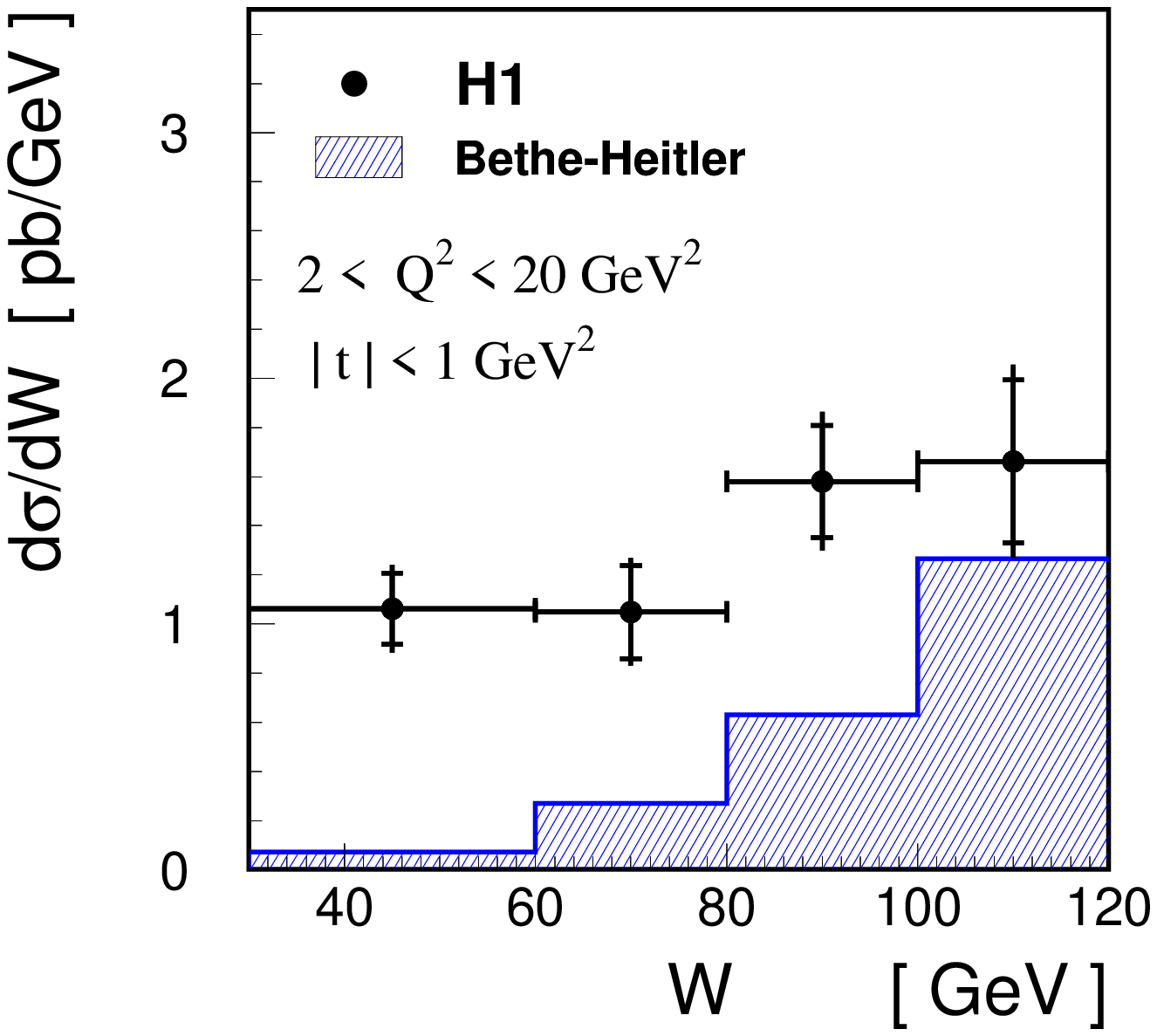,width=0.4\textwidth}
  \vspace{-0.6cm}
  \caption{The measured cross sections of the reaction
  $e^+ p \rightarrow e^+ \gamma p$ differentialy in
  $Q^2$ (left) and $W$ (right) are shown and compared to the pure Bethe-Heitler
  contribution.}
 \label{fig:h_sig1}
  \vspace{-0.5cm}
 \end{center}
 \vspace*{-0.2cm}
\end{figure}

In Fig.~\ref{fig:h_sig1} the total cross sections
of the reaction $e p \rightarrow e \gamma p$ is shown differentialy
in $Q^2$ and in $W$. The data
are compared with the Bethe-Heitler prediction alone (normalised on 
the integrated luminosity). 

\begin{figure}[htbp]
 \begin{center}
  \epsfig{figure=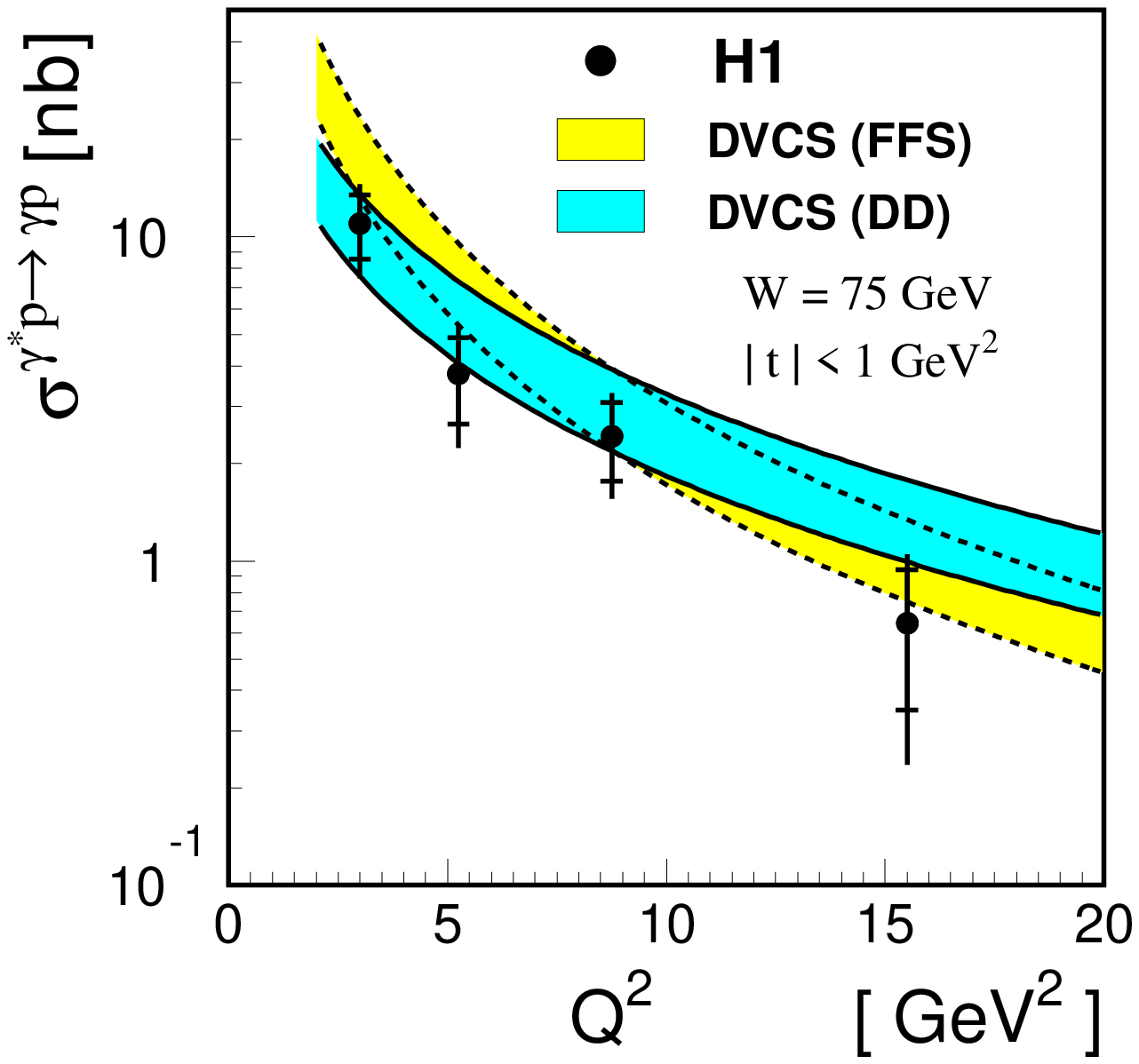,width=0.4\textwidth}
  \epsfig{figure=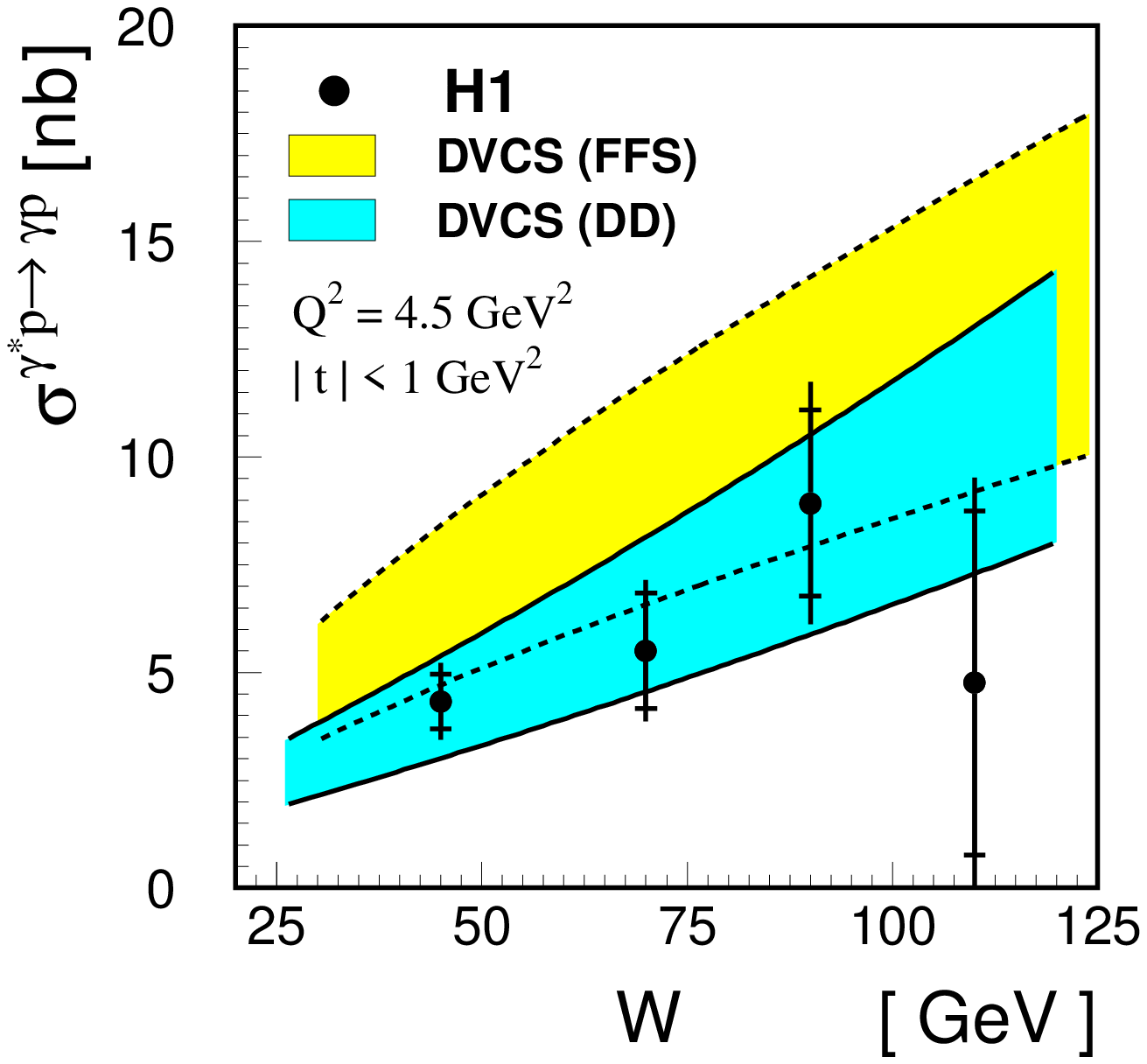,width=0.4\textwidth}
  \vspace{-0.6cm}
  \caption{The measured 
  $\gamma^* p \rightarrow \gamma p$ DVCS cross sections as a function of
  $Q^2$ and $W$. The data are compared to the theoretical predictions
  FFS~\cite{FFS} and DD~\cite{DD}.
  The uncertainty in the theoretical prediction, shown here as a shaded
  band, is dominated
  by the unknown slope of the $t$ dependence of the DVCS part of
  the cross section, assuming $5 < b < 9\, {\rm GeV}^{-2}$.}
 \label{fig:h_sig2}
  \vspace{-0.5cm}
 \end{center}
 \vspace*{-0.2cm}
\end{figure}

The $\gamma^* p \rightarrow \gamma p$ DVCS cross section, derived
subtracting the Bethe-Heitler 
contribution and dividing by the virtual photon flux, is shown as a function 
of $Q^2$ and of $W$ in Fig.~\ref{fig:h_sig2} and compared to FFS and
Donnachie and Dosh (DD)~\cite{DD} predictions.
The description of the data by the calculations is good,
in shape and in absolute normalisation when a $t$ slope is chosen between
5 and 9~GeV$^{-2}$ covering the measured
range for light vector meson production. 
Although the shapes in $Q^2$ and $W$ are similar for the ZEUS
(still preliminary) and H1
measurements, the absolute cross section measurements are not fully
compatible (different $b$ values have to be assumed in the FFS prediction 
to get an agreement). This underlines the importance of the $t$ slope
measurement in the future.

\section{RESULTS DISCUSSION}
\begin{figure}[htbp]
 \begin{picture}(100,100)
  \put(-5,-8){\epsfig{figure=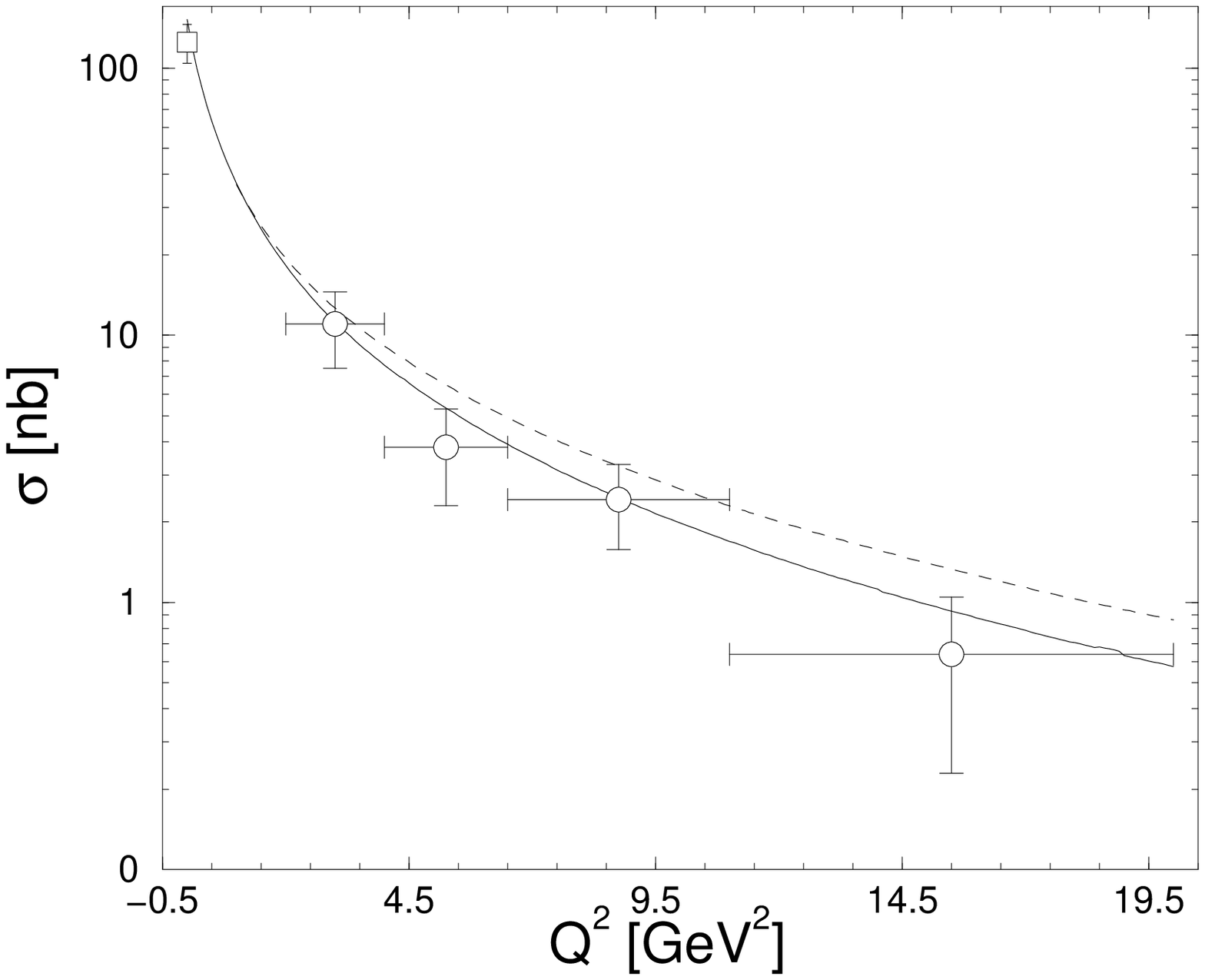,width=0.3\textwidth}}
  \put(133,-2){\epsfig{figure=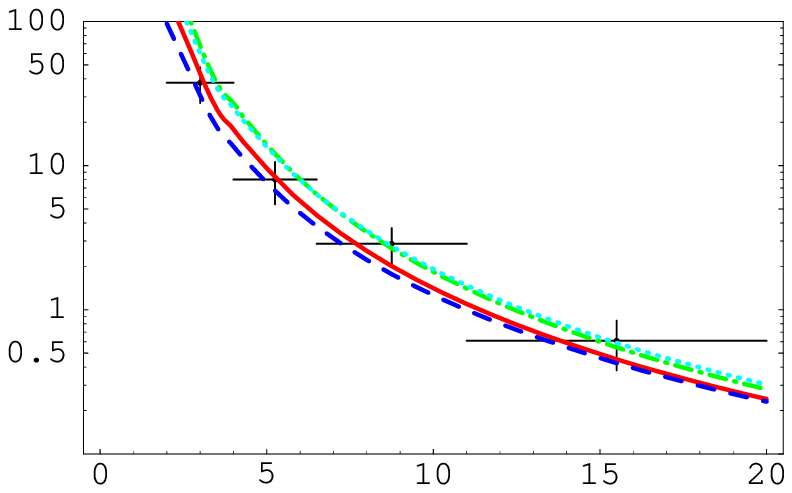,width=0.40\textwidth}}
  \put(305,-11){\epsfig{figure=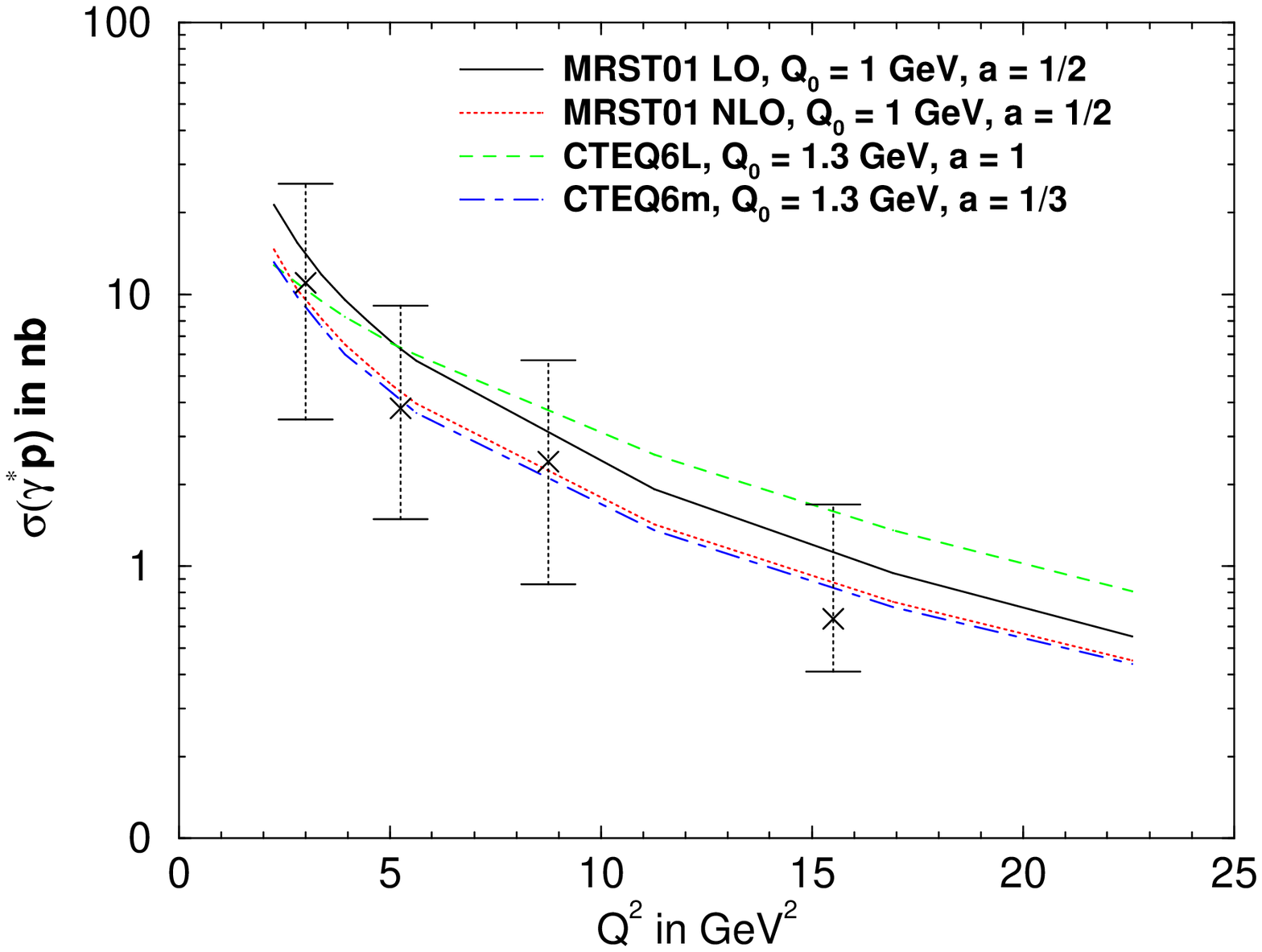,width=0.34\textwidth}}
  \put(20,15){\bf a)}
  \put(160,15){\bf b)}
  \put(330,15){\bf c)}
 \end{picture}
 \caption{The $\gamma^* p \rightarrow \gamma p$ DVCS cross sections
measured by H1 as a function of $Q^2$ 
is compared to different theoretical predictions: \newline
{\bf a)} the dipole models of Forshaw, Kerley and Shaw (full line) 
and of McDermott, Frankfurt, Guzey and Strikman (hashed line)
~\cite{McDermott:2001pt}; \newline
{\bf b)} the LO QCD prediction of Belitski et al.~\cite{Belitsky:2001ns}
going beyond the leading twist but not making a QCD evolution; \newline
{\bf c)} the NLO QCD prediction of Freund and
Mc.\,Dermott~\cite{Freund:2001hd} including QCD evolution of the GPDs.
}
 \label{fig:predictions}
\end{figure}

 In addition to the FFS and DD predictions shown in Fig. ~\ref{fig:h_sig2}, 
soon after publication, the H1 results have been compared to
different predictions by several authors as presented in
Fig.~\ref{fig:predictions} (see caption for the details).
From those confrontations one can conclude several things.
From the dipole model prediction confrontations,
a good agreement is found when one adopts an approach similar to the one
of vector meson electroproduction calculations. It confirms the
diffractive nature of the DVCS process and the validity of the color
dipole models.
The QCD calculation confrontation~\cite{Freund:2001hd} is an important 
success as it consists
in the first diffractive process fully calculated and in good agreement
with the measurement of H1 but also with the helicity asymmetry measurements
of HERMES and CLASS (see~\cite{Freund}). Furthermore the H1 measurement
provides the first constraints on the GPDs (see~\cite{Belitsky:2001ns}
and~\cite{Freund:2001hd} for details).

\section{FUTURE MEASUREMENTS}
 \stepcounter{figure}
 \setcounter{fig:asym}{\value{figure}}
 HERA is presently entering in a high luminosity regime that will
accumulate up to 1~\fbinv\ until 2006, including longitudinal
polarisation of the lepton beam. Additionally, the H1 detector will be
completed by a high acceptance (above  80 \%) proton spectrometer 
(VFPS)~\cite{VFPS}
for $5.10^{-3} < \xpom < 3.10^{-2}$ and 
$|t| \, \lsim \, 0.5$ GeV$^2$,
where $\xpom$ is the proton energy loss.
\\

Within these conditions, charge and helicity asymmetry measurements will
be possible giving access to the Real and the Imaginary parts of the
amplitude separately. 
About 4000 DVCS events are expected with $Q^2 > 8$ GeV~\cite{PRC-ad3}
amoung which about half of them will have a scattered proton measured in
the VFPS.
Results of a first study of the beam charge asymmetry is shown in 
Fig.~\arabic{fig:asym} for a fraction of the full integrated luminosity
(see caption for the details).

\vspace{1.0cm} 
\begin{minipage}[h]{0.4\textwidth}
 {Figure~\arabic{fig:asym}. 
   Beam charge asymmetry simulation for the H1 detector acceptance 
   for different $Q^2$ values and $10<W<120$ GeV,
   assuming an integrated luminosity of 150 \pbinv\ for
   each $e^+$ and $e^-$ sample. Resolution effects are not included.
   }
\end{minipage}

\hspace{0.45\textwidth}
\begin{minipage}[h]{0.5\textwidth}
\vspace{-3.5cm} 
  \epsfig{figure=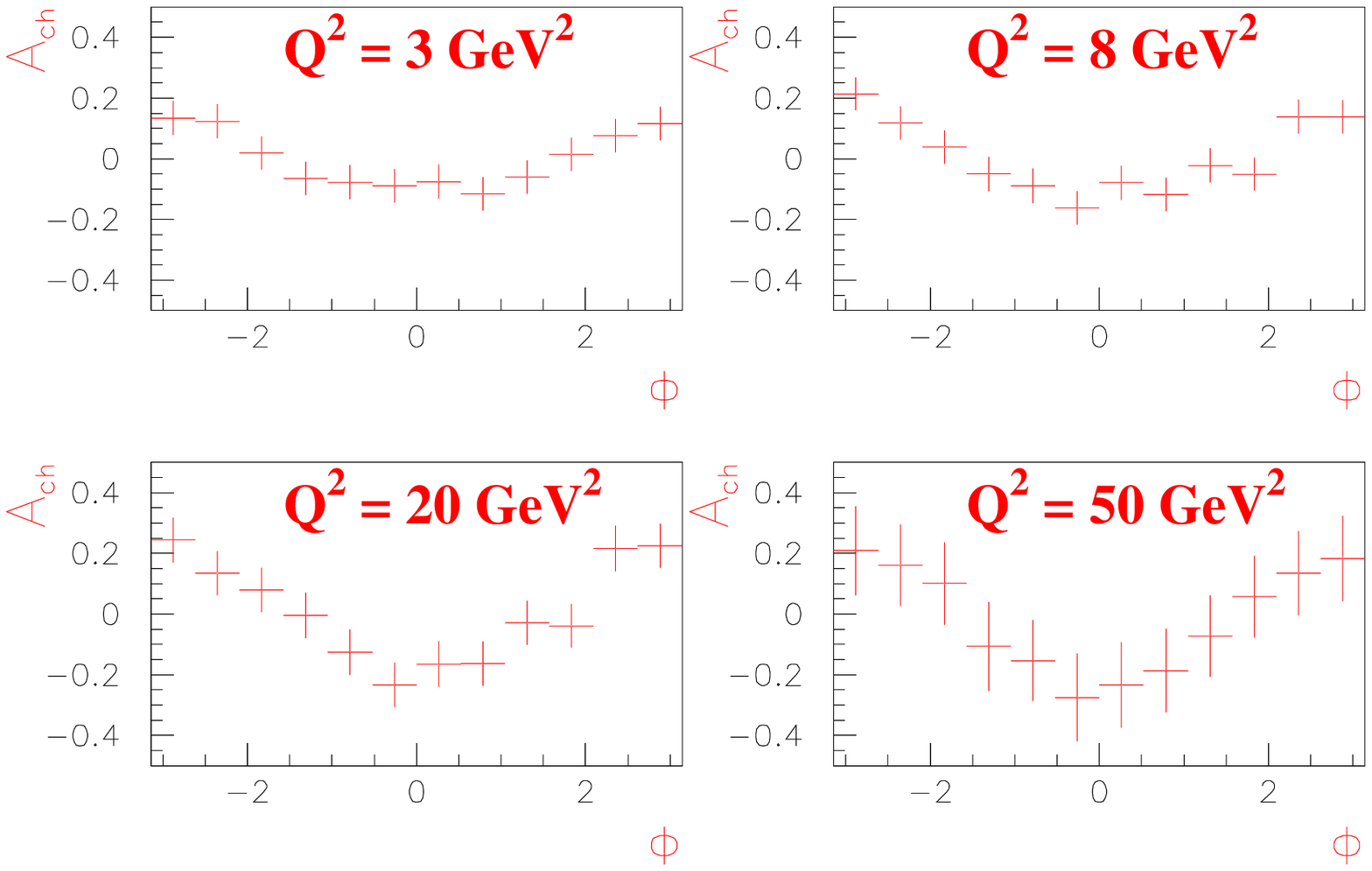,width=0.90\textwidth}
\end{minipage}

\section*{CONCLUSION}

The DVCS process has been observed by the H1 and ZEUS Collaborations.
The first DVCS Cross section measurements derived by H1 and ZEUS 
have been presented.
The experimental results are well described by the pure QCD calculation
at NLO and by several color dipole model predictions.
In the future, HERA will benefit of much larger statistics, and
will yield an access to both the real and the imaginary parts of the
amplitude through beam charge and helicity asymmetry measurements.

\begin{flushright}
{\small
The author is supported by the {\it Fonds National \\
de la Recherche Scientifique} of Belgium.}
\end{flushright}
\end{document}